\documentclass[%
 reprint,
 preprintnumbers,
 nofootinbib,
 amsmath,amssymb,
 aps,
]{revtex4-2}

\usepackage{graphicx}
\usepackage{dcolumn}
\usepackage{bm}
\usepackage{hyperref}
\hypersetup{
	colorlinks=true,
	linkcolor=blue,
	filecolor=magenta,      
	urlcolor=blue,
	citecolor=magenta
}
\usepackage{amsmath,amssymb,mathtools}
\usepackage{graphics,graphicx,tikz,tensor,subcaption}
\usetikzlibrary{arrows,decorations.pathmorphing,patterns}

\begin{document}


\title{Radiation-Reaction and Angular Momentum Loss at $\mathcal{O}(G^4)$}

\author{Carlo Heissenberg$^{\ast}$}
\affiliation{$^{\ast}$Institut de Physique Th\'eorique, CEA Saclay, CNRS, Universit\'e Paris-Saclay, F-91191, Gif-sur-Yvette Cedex, France}



\begin{abstract}
We point out that the odd-in-velocity contribution to the $\mathcal{O}(G^4)$ radiated angular momentum for two-body scattering is determined by the radiation-reaction (RR) term in the one-loop waveform.  This RR term is actually proportional to the tree-level waveform, and this reduces the calculation of the odd-in-velocity contribution to the $\mathcal{O}(G^4)$ angular momentum loss, $J_\text{2rad}$, to two loops, instead of three loops as one would expect by power counting. We exploit this simplification, which follows from unitarity, to obtain a closed-form expression for $J_\text{2rad}$ for generic velocities, which resums all fractional post-Newtonian (PN) corrections to the $\mathcal{O}(G^4)$ angular momentum loss starting at 1.5PN.
\end{abstract}


\maketitle

\section{Introduction} 

The quest for a deeper understanding of the gravitational two-body problem stimulated by the dawn of the gravitational-wave era has recently led to the development of new methods to study the dynamics of compact binaries.
Traditional techniques to tackle this problem are based on the expansion of the classical equations of motion in the weak-field or post-Minkowskian (PM) regime \cite{Peters:1970mx,Kovacs:1977uw,Kovacs:1978eu,Westpfahl:1979gu,Westpfahl:1985tsl} and for small velocities, in the post-Newtonian (PN) regime \cite{Blanchet:2013haa}. For the scattering of two objects with masses $m_1$, $m_2$ at an impact parameter $b$, the PM regime is characterized by  $G m_{1,2}/b\ll1$, while the PN regime holds for $G m_{1,2}/b \sim v^2 \ll1$, with $v$ the relative velocity at infinity.
In recent years, scattering amplitudes have emerged as a powerful tool to recast the PM expansion of gravitational observables in terms of on-shell, gauge-invariant building blocks \cite{Bjerrum-Bohr:2018xdl,Cheung:2018wkq,Kosower:2018adc,KoemansCollado:2019ggb,Cristofoli:2021vyo,Cristofoli:2021jas,DiVecchia:2023frv}. Amplitudes offer an independent way of organizing such calculations, which can serve to more easily identify new structures, simplify the computation of higher-order contributions and advance the precision frontier.  

An example of such a simplification was the inclusion of radiation-reaction (RR) in the $\mathcal{O}(G^3)$ deflection, which is the odd-in-velocity, hence half-odd PN, correction to the 3PM deflection angle \cite{DiVecchia:2020ymx,Damour:2020tta,DiVecchia:2021ndb,DiVecchia:2021bdo,Herrmann:2021tct,Heissenberg:2021tzo}. While the complete $\mathcal{O}(G^3)$ result requires computing the two-loop amplitude in Fig.~\ref{fig:2loop22} \cite{Bern:2019nnu,Bern:2019crd,Kalin:2020fhe,Bjerrum-Bohr:2021vuf,Bjerrum-Bohr:2021din,Damgaard:2021ipf,Brandhuber:2021eyq,Jakobsen:2022psy,Kalin:2022hph}, the RR contribution to its real part is determined, via unitarity and analyticity, by the infrared (IR) divergence in the imaginary part due to  three-particle cut in Fig.~\ref{fig:3pc} \cite{DiVecchia:2021ndb}. In this way, the RR contribution at $\mathcal{O}(G^3)$ takes the form of a much simpler one-loop integral, Fig.~\ref{fig:22simplified}, times an elementary function.

\begin{figure}
	\hspace{-15pt}
	\begin{subfigure}[t]{0.12\textwidth}
		\begin{tikzpicture}
			\draw [ultra thick, blue] (-.8,.8) -- (0,0);
			\draw [ultra thick, blue] (.8,.8) -- (0,0);
			\draw [ultra thick, green!70!black] (-.8,-.8) -- (0,0);
			\draw [ultra thick, green!70!black] (.8,-.8) -- (0,0);
			\filldraw[black!20!white, thick] (0,0) ellipse (0.5 and 0.5);
			\draw[thick] (0,0) ellipse (0.5 and 0.5);
			\filldraw[white, thick] (0,.2) ellipse (0.25 and 0.15);
			\draw[thick] (0,.2) ellipse (0.25 and 0.15);
			\filldraw[white, thick] (0,-.2) ellipse (0.25 and 0.15);
			\draw[thick] (0,-.2) ellipse (0.25 and 0.15);
		\end{tikzpicture}
		\caption{\label{fig:2loop22}}
	\end{subfigure}%
	~
	\begin{subfigure}[t]{0.16\textwidth}
		\begin{tikzpicture}
			\draw[red,decorate,decoration={coil,,aspect=0,segment length=1.5mm,amplitude=.5mm,pre length=0pt,post length=0pt}]  (0,0) -- (1,0);
			\draw [ultra thick, blue] (-.8,.6) -- (0,0);
			\draw [ultra thick, blue] (0,0) .. controls (0,.5) and (1,.5) .. (1,0);
			\draw [ultra thick, blue] (1,0) -- (1.8,.6);
			\draw [ultra thick, green!70!black] (-.8,-.6) -- (0,0);
			\draw [ultra thick, green!70!black] (0,0) .. controls (0,-.5) and (1,-.5) .. (1,0);
			\draw [ultra thick, green!70!black] (1,0) -- (1.8,-.6);
			\filldraw[black!20!white, thick] (0,0) ellipse (0.3 and 0.3);
			\draw[thick] (0,0) ellipse (0.3 and 0.3);
			\filldraw[black!20!white, thick] (1,0) ellipse (0.3 and 0.3);
			\draw[thick] (1,0) ellipse (0.3 and 0.3);
		\end{tikzpicture}
		\caption{\label{fig:3pc}}
	\end{subfigure}%
	~
	\begin{subfigure}[t]{0.16\textwidth}
			\begin{tikzpicture}
			\draw [ultra thick, blue] (-.8,.6) -- (0,0);
			\draw [ultra thick, blue] (0,0) .. controls (0,.5) and (1,.5) .. (1,0);
			\draw [ultra thick, blue] (1,0) -- (1.8,.6);
			\draw [ultra thick, green!70!black] (-.8,-.6) -- (0,0);
			\draw [ultra thick, green!70!black] (0,0) .. controls (0,-.5) and (1,-.5) .. (1,0);
			\draw [ultra thick, green!70!black] (1,0) -- (1.8,-.6);
			\filldraw[black!20!white, thick] (0,0) ellipse (0.3 and 0.3);
			\draw[thick] (0,0) ellipse (0.3 and 0.3);
			\filldraw[black!20!white, thick] (1,0) ellipse (0.3 and 0.3);
			\draw[thick] (1,0) ellipse (0.3 and 0.3);
		\end{tikzpicture}
		\caption{\label{fig:22simplified}}
	\end{subfigure}%
	\caption{Amplitudes relevant for the classical impulse. Thick solid lines depict massive particles and a wiggly line represents a graviton. Exposed lines are on-shell.}
\end{figure}

Using the classical limit of scattering amplitudes and worldline approaches that efficiently solve the classical equations of motion \cite{Kalin:2020mvi,Mogull:2020sak,Dlapa:2023hsl}, a full calculation of the $\mathcal{O}(G^4)$ impulses (three loops) was achieved \cite{Bern:2021dqo,Dlapa:2021npj,Bern:2021yeh,Dlapa:2021vgp,Dlapa:2022lmu,Damgaard:2023ttc} and progress is underway at $\mathcal{O}(G^5)$ (four loops) \cite{Driesse:2024xad,Bern:2024adl,Driesse:2024feo}. 
Such methods also mesh well with complementary approximation principles including the self-force  expansion, in which calculations are organized in powers of the mass ratio of the binary \cite{Cheung:2023lnj,Kosmopoulos:2023bwc,Cheung:2024byb,Damgaard:2024fqj,Mougiakakos:2024nku,Mougiakakos:2024lif} (see also \cite{Adamo:2022qci,Adamo:2023cfp,Fucito:2024wlg}).

Another interesting observable is of course the gravitational waveform, which is the dynamical metric fluctuation, $g_{\mu\nu}-\eta_{\mu\nu}\sim \frac{4G}{r}\,w_{\mu\nu}$, measured by a detector placed at a large distance $r$ from the sources. The $\mathcal{O}(G)$  contribution to $w_{\mu\nu}$, obtained in \cite{Peters:1970mx,Kovacs:1977uw,Kovacs:1978eu} and more recently streamlined in \cite{Jakobsen:2021smu,Mougiakakos:2021ckm}, is given by a tree-level amplitude (Fig.~\ref{fig:tree23}) involving one graviton emission \cite{Goldberger:2016iau,Luna:2017dtq} and entails integer PN corrections to the waveform multipoles.

\begin{figure}[b!]
	\hspace{-10pt}
		\begin{subfigure}[t]{0.13\textwidth}
		\begin{tikzpicture}
			\draw[red,decorate,decoration={coil,,aspect=0,segment length=1.5mm,amplitude=.5mm,pre length=0pt,post length=0pt}] (0,0)  -- (1.13,0);
			\draw [ultra thick, blue] (-.8,.8) -- (0,0);
			\draw [ultra thick, blue] (.8,.8) -- (0,0);
			\draw [ultra thick, green!70!black] (-.8,-.8) -- (0,0);
			\draw [ultra thick, green!70!black] (.8,-.8) -- (0,0);
			\filldraw[black!20!white, thick] (0,0) ellipse (0.5 and 0.5);
			\draw[thick] (0,0) ellipse (0.5 and 0.5);
			\filldraw[white, thick] (0,0) ellipse (0.25 and 0.25);
			\draw[thick] (0,0) ellipse (0.25 and 0.25);
		\end{tikzpicture}
		\caption{\label{fig:1loop23}}
		\end{subfigure}
		~
		\begin{subfigure}[t]{0.17\textwidth}
			\begin{tikzpicture}
				\draw[red,decorate,decoration={coil,,aspect=0,segment length=1.5mm,amplitude=.5mm,pre length=0pt,post length=0pt}] (0,0) .. controls (.5,0) and (1,0.3) .. (1,.6);
				\draw[red,decorate,decoration={coil,,aspect=0,segment length=1.5mm,amplitude=.5mm,pre length=0pt,post length=0pt}] (2,0) .. controls (1.5,0) and (1.2,0.3) .. (1.2,.6);
				\draw [ultra thick, blue] (-.8,.6) -- (0,0);
				\draw [ultra thick, blue] (0,0) .. controls (0,.5) and (.5,.6) .. (1,.6);
				\draw [ultra thick, blue] (1,.6) -- (2,.6);
				\draw [ultra thick, green!70!black] (-.8,-.6) -- (0,0);
				\draw [ultra thick, green!70!black] (0,0) .. controls (0,-.5) and (.5,-.6) .. (1,-.6);
				\draw [ultra thick, green!70!black] (1,-.6) -- (2,-.6);
				\filldraw[black!20!white, thick] (0,0) ellipse (0.3 and 0.3);
				\draw[thick] (0,0) ellipse (0.3 and 0.3);
				\filldraw[black!20!white, thick] (1,.6) ellipse (0.3 and 0.3);
				\draw[thick] (1,.6) ellipse (0.3 and 0.3);
			\end{tikzpicture}
			\caption{\label{fig:Compton}}
		\end{subfigure}
		~
		\begin{subfigure}[t]{0.13\textwidth}
			\begin{tikzpicture}
				\draw[red,decorate,decoration={coil,,aspect=0,segment length=1.5mm,amplitude=.5mm,pre length=0pt,post length=0pt}] (0,0)  -- (.83,0);
				\draw [ultra thick, blue] (-.8,.6) -- (0,0);
				\draw [ultra thick, blue] (.8,.6) -- (0,0);
				\draw [ultra thick, green!70!black] (-.8,-.6) -- (0,0);
				\draw [ultra thick, green!70!black] (.8,-.6) -- (0,0);
				\filldraw[black!20!white, thick] (0,0) ellipse (0.3 and 0.3);
				\draw[thick] (0,0) ellipse (0.3 and 0.3);
			\end{tikzpicture}
			\caption{\label{fig:tree23}}
		\end{subfigure}%
\caption{Amplitudes relevant for the waveform.}
\end{figure}

The next order, $\mathcal{O}(G^2)$, can be expressed in terms of the one-loop amplitude in Fig.~\ref{fig:1loop23}  \cite{Brandhuber:2023hhy,Herderschee:2023fxh,Elkhidir:2023dco,Georgoudis:2023lgf} (see also \cite{Caron-Huot:2023vxl,Georgoudis:2023ozp,Georgoudis:2023eke}).
Once again, the real part of this amplitude contains a RR piece that is induced via unitarity and analyticity by the IR divergence in the imaginary part due to the Compton or rescattering cuts in Fig.~\ref{fig:Compton}. 
This RR contribution is simply proportional to the tree-level amplitude in Fig.~\ref{fig:tree23}.
At this order, in addition to integer PN, or ``instantaneous'', contributions, the multipolar waveform thus also includes half-odd PN terms which are entirely captured by the simple RR contribution and by the  Compton cuts \cite{Bini:2023fiz,Georgoudis:2023eke,Georgoudis:2024pdz,Bini:2024rsy}, which encode the so-called tail effect whereby gravitational radiation scatters off the curvature produced by the massive objects.

Knowledge of the gravitational waveform also allows one to study the energy and angular momentum lost by the binary due to the interaction with the gravitational field, as achieved in \cite{Herrmann:2021lqe,Riva:2021vnj,Manohar:2022dea,DiVecchia:2022piu} at $\mathcal{O}(G^3)$. One of the key steps of such calculations  consists in recasting the resulting phase-space integration in terms of the two-loop cut in Fig.~\ref{fig:3pc} (the ``square'' of Fig.~\ref{fig:tree23}) via reverse unitarity \cite{Anastasiou:2002yz,Anastasiou:2002qz,Anastasiou:2003yy,Anastasiou:2015yha}. The $\mathcal{O}(G^4)$ energy loss can be deduced from the $\mathcal{O}(G^4)$ impulses in \cite{Dlapa:2022lmu,Damgaard:2023ttc}, while the analysis of the $\mathcal{O}(G^4)$ angular momentum loss was initiated in \cite{Heissenberg:2024umh} by obtaining a closed-form expression for the so-called static contribution \cite{Damour:2020tta,Manohar:2022dea,DiVecchia:2022owy,Bini:2022wrq,Riva:2023xxm, Heissenberg:2024umh,Biswas:2024ept}. 
In \cite{Heissenberg:2024umh}, the total angular momentum loss was eventually analyzed only in the PN limit, thus recovering the results in \cite{Bini:2021gat,Bini:2022enm} for the 1.5PN and 2.5PN center-of-mass angular momentum loss. 

In this work, we clarify the connection between the building blocks of the one-loop waveform and the $\mathcal{O}(G^4)$ radiated linear and angular momentum. We provide a parity argument based on the unitarity properties of the one-loop waveform kernel showing that all fractional PN  corrections to the $\mathcal{O}(G^4)$ energy $E_\text{2rad}$ and angular momentum $J_\text{2rad}$ losses are independent of the Compton cuts and can be entirely reduced to the RR contribution of the one-loop waveform. This RR contribution is proportional to the tree-level waveform, and this reduces a naively three-loop calculation to a two-loop one, which we perform using the tools developed in \cite{Herrmann:2021lqe,DiVecchia:2022piu}. In this way, we derive simple closed-form expressions for $E_\text{2rad}$ and $J_\text{2rad}$ valid for generic velocities. The former serves as a cross-check of the results in \cite{Dlapa:2022lmu,Damgaard:2023ttc}, while the latter constitutes a new result that resums all half-odd PN corrections to the $\mathcal{O}(G^4)$ angular momentum loss.

\section{Structure of the gravitational waveform up to one loop}

We start by briefly reviewing the structure of the waveform ``kernel''  (working with $\eta_{\mu\nu}=\text{diag}(- + + +)$),
\begin{equation}\label{eq:kernelW}
	W^{\mu\nu} = W_0^{\mu\nu} + W_1^{\mu\nu} + \cdots
\end{equation}
where $W_0$ is the tree-level, $\mathcal{O}(G^{3/2})$,  and $W_1$ the one-loop,  $\mathcal{O}(G^{5/2})$, contribution. This is the classical object whose Fourier transform \eqref{eq:tildeW} to impact parameter $b^\alpha=b_1^\alpha-b_2^\alpha$
gives the gravitational waveform
\begin{equation}\label{eq:theWaveform}
	g_{\mu\nu}-\eta_{\mu\nu} \sim \frac{4G}{r}\int_0^\infty \frac{d\omega}{2\pi}\,\frac{\tilde{W}_{\mu\nu}(\omega\, n)}{\sqrt{8\pi G}}\,e^{-i\omega U}+\text{c.c.}
\end{equation}
at retarded time $U$ and angles $n^\mu$.
See Refs.~\cite{Georgoudis:2023eke,Georgoudis:2024pdz,Bini:2024rsy} for further details. In this work, we focus on the scattering of minimally-coupled massive scalar objects.

The tree-level piece is simply the classical limit of the tree-level amplitude  
$W_0 = \mathcal{A}_0$ \cite{Goldberger:2016iau,Luna:2017dtq,DiVecchia:2021bdo}, see Fig.~\ref{fig:tree23}. Upon Fourier transform, $\tilde{\mathcal{A}}_0$ thus provides the leading PM waveform \cite{Kovacs:1978eu,Goldberger:2016iau,Jakobsen:2021smu,Mougiakakos:2021ckm}. When expanded for small velocity, $\tilde{\mathcal{A}}_0$ starts with a Newtonian (0PN) contribution captured by the Einstein quadrupole formula, and then yields (relative) integer PN corrections to each multipole.

The one-loop kernel ${W}_{1}$ is obtained from the one-loop amplitude $\mathcal{A}_1$ \cite{Brandhuber:2023hhy,Herderschee:2023fxh,Elkhidir:2023dco,Georgoudis:2023lgf}, which involves a real part plus unitarity cuts,
\begin{equation}
	\mathcal{A}_1 ^{\mu\nu}= \mathcal{B}_1^{\mu\nu} + \frac{i}{2}(s^{\mu\nu}+s'^{\mu\nu}) + \frac{i}{2} (c_1^{\mu\nu}+c_2^{\mu\nu})\,,
\end{equation}
by dropping the two-massive-particle cuts $s$ and $s'$ \cite{Georgoudis:2023eke},
\begin{equation}\label{eq:kernelW1}
	{W}_1^{\mu\nu} = \mathcal{B}^{\mu\nu}_1 + \frac{i}{2} (c^{\mu\nu}_1+c^{\mu\nu}_2)\,,
\end{equation}
keeping the real part $\mathcal{B}_1$ and the Compton cuts $c_1$ (see Fig.~\ref{fig:Compton}) and $c_2$. 
The simple form of the one-loop kernel \eqref{eq:kernelW1} obtains when the waveform is expressed in terms of the \emph{average} velocities of the scattering objects, $u_1^\alpha$, $u_2^\alpha$, and of the \emph{eikonal} impact parameter $b^\alpha$ orthogonal to them \cite{Georgoudis:2023eke} (see also \cite{Caron-Huot:2023vxl,Bini:2023fiz,Aoude:2023dui,Georgoudis:2024pdz,Bini:2024rsy}). These are also the standard reference vectors employed in the PN literature 
(see e.g.~\cite{Bini:2021gat}) and, at this order, they differ from the initial ones, $v_1^\alpha$, $v_2^\alpha$, $b_{J}^\alpha$, by an $\mathcal{O}(G)$ 
rotation.\footnote{This distinction is completely irrelevant for the main results for the 2rad contributions in \eqref{eq:2radP}, \eqref{eq:2radJ}. Indeed,  a rotation of the reference vectors by $\frac{1}{2}\Theta_\text{1PM}$, which has an integer PN expansion, would only induce a mixing between the 1rad contributions \eqref{eq:1radPK}, \eqref{eq:1radJO} and their $\mathcal{O}(G^3)$ counterparts \eqref{eq:PLO}, \eqref{eq:JLO}. More precisely, by \eqref{eq:bPG3=0}, \eqref{eq:uJuG3=0}, it would induce nonzero feed-down terms in the right-hand sides of \eqref{eq:1rad0components}. For instance, one can check, using the results in Ref.~\cite{Dlapa:2022lmu}, that $b\cdot( \boldsymbol{P}_{\mathcal{O}(G^3)}+ \boldsymbol{P}_{\text{1rad}})=0$, while $b_J\cdot( \boldsymbol{P}_{\mathcal{O}(G^3)}+ \boldsymbol{P}_{\text{1rad}})\neq0$.}
We define the invariant $\sigma = -v_1\cdot v_2$, which is the relative Lorentz factor between the two objects, and the frequencies $\omega_1 = -v_1 \cdot k$, $\omega_2 = -v_2\cdot k$. The center-of-mass energy $E = \sqrt{m_1^2+2 m_1 m_2 \sigma + m_2^2}$ and frequency $\omega$ obey 
	\begin{equation}
		E \omega = m_1 \omega_1 + m_2 \omega_2\,.
	\end{equation} 

The Compton cuts $c_1$, $c_2$ take into account the rescattering of gravitational radiation against the curvature sourced by the binary system. They are infrared divergent \cite{Weinberg:1965nx}, and this divergence can be resummed into a phase factor, $W =
e^{-i GE\omega/\epsilon}\,
W_\text{reg}$, with $\epsilon = \frac{1}{2}(4-D)$ the dimensional regulator
and 
\begin{equation}\label{eq:kernelWreg}
		W_\text{reg}^{\mu\nu}
	=
	\mathcal{A}_0^{\mu\nu} + \mathcal{B}_1^{\mu\nu} + \frac{i}{2}\, \mathcal{C}^{\mu\nu}  + \cdots
\end{equation}
The divergence can thus be reabsorbed by redefining the origin of retarded time in \eqref{eq:theWaveform} \cite{Goldberger:2009qd,Porto:2012as}, and the regulated Compton cuts $\mathcal{C}$ 
contain the logarithm of an unspecified energy scale $\mu_{\text{IR}}$, which amounts to performing further finite time translations by $2G E \log\mu_{\text{IR}}$. The remainder $\tilde{C}^\text{reg}$ 
(see \eqref{eq:ComptonCutsExplicit})
contributes half-odd corrections to the multipolar waveform starting at 1.5PN order.

The real part $\mathcal{B}_1$ is further composed of an ``Odd'' part, which is proportional to the tree-level amplitude, and an instantaneous ``Even'' part,
$\mathcal{B}_{1} = \mathcal{B}_{1O} + \mathcal{B}_{1E}$.
We  further split $ \mathcal{B}_{1O} $ as
$\mathcal{B}_{1O} = \mathcal{B}_{1O}^{(i)} + \mathcal{B}_{1O}^{(h)}$
with
\begin{subequations}
	\begin{align}\label{eq:radreac(i)}
		\mathcal{B}_{1O}^{(i){\mu\nu}}
		&=-\frac{\sigma(\sigma^2-3/2)}{(\sigma^2-1)^{3/2}}\, \pi G E \omega \, \mathcal{A}^{\mu\nu}_0\,,\\
		\label{eq:radreac(h)}
		\mathcal{B}_{1O}^{(h){\mu\nu}}
		&= \pi G E \omega \, \mathcal{A}^{\mu\nu}_0\,.
	\end{align}
\end{subequations}
While $\tilde{\mathcal{B}}_{1O}^{(i)}$, $\tilde{\mathcal{B}}_{1E}$ only contribute integer PN corrections to the multipolar waveform, which start at 0PN and 1PN orders, $\tilde{\mathcal{B}}_{1O}^{(h)}$ only contributes half-odd corrections starting at 1.5PN order and identifies the RR contribution.

The half-odd PN contributions are thus completely captured by the RR part \eqref{eq:radreac(h)} and by the (regulated) Compton cuts.
From here on, we shall drop the subscript ``reg'' and always work with the regulated waveform.

\section{Radiated linear and angular momentum}

The radiated energy-momentum and angular momentum-mass dipole moment are  given by the following expressions in terms of the waveform $\tilde{W}$ obeying $k^\mu \tilde{W}_{\mu\nu}=0$,
\begin{equation}\label{eq:generalPJ}
	\boldsymbol{P}^{\alpha} 
	=
	\int_k \boldsymbol{K}^\alpha[\tilde W, \tilde W]\,,
	\quad
	\boldsymbol{J}^{\alpha\beta}
	=
	-i \int_k \boldsymbol{O}^{\alpha\beta}[\tilde W, \tilde W]\,,
\end{equation}
where $\int_k$ is a shorthand for the phase-space integral \eqref{eq:intk}
and the integrands $\boldsymbol{K}_\alpha$, $\boldsymbol{O}_{\alpha\beta}$ are given by
 \cite{Herrmann:2021lqe}
\begin{equation}\label{eq:Kdef}
	\boldsymbol{K}_\alpha[\tilde{X}, \tilde{Y}]
	=
	D^{\mu\nu,\rho\sigma}
	k_\alpha\,
	\tilde{X}_{\mu\nu}^{\ast}
	\tilde{Y}_{\rho\sigma}
\end{equation}
with $D^{\mu\nu,\rho\sigma} = \eta^{\mu\rho}
\eta^{\nu\sigma}
-
\eta^{\mu\nu}
\eta^{\rho\sigma}/(D-2)$
and
\cite{Manohar:2022dea,DiVecchia:2022owy}
\begin{equation}\label{eq:Odef}
\boldsymbol{O}_{\alpha\beta}[\tilde{X}, \tilde{Y}]
=
D^{\mu\nu,\rho\sigma} 
	\tilde{X}_{\mu\nu}^{\ast}
	k_{[\alpha}^{\phantom{\mu}}
	\frac{ \overset{\leftrightarrow}{\partial}}{\partial k^{\beta]}} \tilde{Y}_{\rho\sigma}
	+2
	\tilde{X}_{\mu[\alpha}^{\ast}
	\tilde{Y}_{\beta]}^{\mu}\,.
\end{equation}
Here $A_{[\alpha} B_{\beta]} = A_\alpha B_\beta-A_{\beta}B_{\alpha}$ and $f\overset{\leftrightarrow}{\partial}g = \frac{1}{2}(f\,\partial g-g\,\partial f)$. Note for later convenience that
\begin{subequations}
\begin{align}\label{eq:parity}
	\boldsymbol{K}^{\alpha}[\tilde X^\ast,\tilde Y^\ast]
	&=
	\boldsymbol{K}^{\alpha}[\tilde X,\tilde Y]^\ast \,\,= + \boldsymbol{K}^{\alpha}[\tilde Y,\tilde X]\,,
	\\
	\label{eq:parityJ}
	\boldsymbol{O}^{\alpha\beta}[\tilde X^\ast,\tilde Y^\ast]
	&=
	\boldsymbol{O}^{\alpha\beta}[\tilde X,\tilde Y]^\ast = - \boldsymbol{O}^{\alpha\beta}[\tilde Y,\tilde X]\,.
\end{align}
\end{subequations}

We recall that $\boldsymbol{P}^\alpha$ and $\boldsymbol{J}^{\alpha\beta}$ are Lorentz tensors, and that $\boldsymbol{P}^\alpha$  is translation-invariant, while
\begin{equation}\label{eq:translation}
	\boldsymbol{J}^{\alpha\beta} \mapsto \boldsymbol{J}^{\alpha\beta} + a^{[\alpha} \boldsymbol{P}^{\beta]}
\end{equation}
under a translation by $a^\alpha$.

The notation for the integrands \eqref{eq:Kdef}, \eqref{eq:Odef} is introduced to more easily discuss the various contributions arising when inserting the PM expanded waveform \eqref{eq:kernelWreg} into the expressions \eqref{eq:generalPJ}, which gives
\begin{subequations}
\begin{align}
	\boldsymbol{P}^\alpha &= \boldsymbol{P}_{\mathcal{O}(G^3)}^\alpha +  \boldsymbol{P}_{\mathcal{O}(G^4)}^\alpha + \cdots
	\\
	\boldsymbol{J}^{\alpha\beta} &= \boldsymbol{J}_{\mathcal{O}(G^3)}^{\alpha\beta} +  \boldsymbol{J}_{\mathcal{O}(G^4)}^{\alpha \beta} + \cdots
\end{align}
\end{subequations}
We discuss their properties, choosing the translation frame $b_1^\alpha = b^\alpha$, $b_2^\alpha=0$ without loss of generality.

The leading-order contribution to the linear momentum is given by retaining only the tree-level contribution to Eq.~\eqref{eq:kernelWreg},
\begin{equation}\label{eq:PLO}
	\boldsymbol{P}_{\mathcal{O}(G^3)}^{\alpha} 
	=
	\int_k \boldsymbol{K}^{\alpha}_0\,,\qquad
	\boldsymbol{K}^\alpha_0 = \boldsymbol{K}^\alpha[\tilde{\mathcal{A}}_0, \tilde{\mathcal{A}}_0]\,.
\end{equation}
Let us note that
$\tilde{\mathcal{A}}_0^\ast = \tilde{\mathcal{A}}_0\big|_{b\mapsto -b}$,
which follows from the reality of the  momentum-space tree-level amplitude $\mathcal{A}_0$ entering the Fourier transform  \eqref{eq:tildeW}. On the other hand, $\boldsymbol{K}^{\alpha\ast}_0 = \boldsymbol{K}^\alpha_0$ is real, so 
\begin{equation}
	\boldsymbol{K}^\alpha_0 = \boldsymbol{K}^\alpha_0\big|_{b\to -b}\,.
\end{equation} 
The coefficients of the form-factor decomposition, 
\begin{equation}
	\boldsymbol{K}^\alpha_0 = f_{u_1}\,\check{u}_1^\alpha + f_{u_2}\,\check{u}_2^\alpha + f_{b}\, b^\alpha + f_k\,k^\alpha\,,
\end{equation}
with\footnote{Explicitly, $\check{u}_1^\alpha = (\sigma u_2^\alpha-u_1^\alpha)/(\sigma^2-1)$, $\check{u}_2^\alpha = (\sigma u_1^\alpha-u_2^\alpha)/(\sigma^2-1)$ up to $\mathcal{O}(G^2)$ corrections.} 
$\check{u}_{i}\cdot u_j = -\delta_{ij}$,
are thus real functions of the invariant products with definite parity under $b\mapsto -b$ (since only the invariant product $b\cdot k$ transforms, we only highlight this argument)
\begin{subequations}
	\begin{align}
		f_{u_{1,2}}(-b\cdot k) &= +  f_{u_{1,2}}(b\cdot k)\,,\\
		f_{b}(-b\cdot k) &= -  f_{b}(b\cdot k)\,,\\
		f_{k}(-b\cdot k) &= +  f_{k}(b\cdot k)\,.
	\end{align}
\end{subequations}
Therefore, the integrand 
\begin{equation}
	b\cdot \boldsymbol{K}_0 = f_b\,b^2 + f_k\, b\cdot k
\end{equation}
appearing in $b\cdot \boldsymbol{P}_{\mathcal{O}(G^3)}$ is
\emph{odd} under the change of 
variable
$b\cdot k\mapsto -b\cdot k$, while the phase-space measure is even,\footnote{We note that $-u_1 \cdot k$, $-u_2\cdot k$, $b\cdot k$ and $k_o^\mu$ with $k_o\cdot u_{1,2}=0=k_o\cdot b$ are $D$ independent variables for the phase-space integration over $k^\mu = -u_1 \cdot k \,\check{u}_1^\mu- u_2 \cdot k \,\check{u}_2^\mu + b\cdot k\,b^\mu /b^2+k^\mu_o$. Furthermore, $k^2=P(u_1\cdot k, u_2\cdot k)+(b\cdot k)^2/b^2 + k_o^2$, where $P$ is independent of $b\cdot k$, so that $k^2$ is even under $b\cdot k\mapsto -b\cdot k$.} 
and therefore we recover \cite{Herrmann:2021lqe}
\begin{equation}\label{eq:bPG3=0}
	b\cdot \boldsymbol{P}_{\mathcal{O}(G^3)} = 0.
\end{equation}
Conversely, the integrands for $-u_{1,2}\cdot \boldsymbol{P}_{\mathcal{O}(G^3)}$ are even and  these components are indeed nontrivial \cite{Herrmann:2021lqe}.

Turning to the leading contribution to the radiated angular momentum,
\begin{equation}\label{eq:JLO}
	\boldsymbol{J}_{\mathcal{O}(G^3)}^{\alpha\beta} 
	=-i
	\int_k \boldsymbol{O}_0^{\alpha\beta}\,,\qquad
	\boldsymbol{O}^{\alpha\beta}_0 = \boldsymbol{O}^{\alpha\beta}[\tilde{\mathcal{A}}_0, \tilde{\mathcal{A}}_0]\,,
\end{equation}
instead we note that $\boldsymbol{O}^{\alpha\beta\ast}_0 = - \boldsymbol{O}^{\alpha\beta}_0$ is imaginary, so
\begin{equation}
	\boldsymbol{O}^{\alpha\beta}_0  = - \boldsymbol{O}^{\alpha\beta}_0 \big|_{b\mapsto-b}\,.
\end{equation}
Therefore
\begin{equation}
	\begin{split}
	\boldsymbol{O}^{\alpha\beta}_0  
	&= i \Big(
	f_{u_1u_2}\,\check{u}_1^{[\alpha} \check{u}_2^{\beta]}
	+
	f_{u_1 b}\,\check{u}_1^{[\alpha} b^{\beta]}
	+
	f_{u_2 b}\,\check{u}_2^{[\alpha} b^{\beta]}
	\\
	&+
	f_{u_1 k}\,\check{u}_1^{[\alpha} k^{\beta]}
	+
	f_{u_2 k}\,\check{u}_2^{[\alpha} k^{\beta]}
	+
	f_{b k}\,b^{[\alpha} k^{\beta]}
	\Big)
	\end{split}
\end{equation}
where the coefficients are real functions transforming as
\begin{equation}
	\begin{split}
		f_{u_1 u_2}(-b\cdot k) &= -  f_{u_1 u_2}(b\cdot k)\,,\\
		f_{u_{1,2} b}(-b\cdot k) &= +  f_{u_{1,2} b}(b\cdot k)\,,\\
		f_{u_{1,2} k}(-b\cdot k) &= -  f_{u_{1,2} k}(b\cdot k)\,,\\
		f_{b k}(-b\cdot k) &= + f_{b k}(b\cdot k)\,.
	\end{split}
\end{equation}
From this, it is clear that 
\begin{equation}\label{eq:uJuG3=0}
u_1\cdot \boldsymbol{J}_{\mathcal{O}(G^3)}\cdot u_2=0\,,
\end{equation} 
because the integrand for this component is odd under $b\cdot k\mapsto-b\cdot k$. This agrees with the  results in \cite{Manohar:2022dea,DiVecchia:2022piu}.

Moving to the next order in $G$, we need to consider the interference terms between the tree-level and the two-loop ingredients of the waveform \eqref{eq:kernelWreg}. Following the terminology employed in \cite{Dlapa:2022lmu}, we can arrange the resulting terms as follows, $\boldsymbol{P}_{\mathcal{O}(G^4)}^\alpha = \boldsymbol{P}_\text{1rad}^\alpha+\boldsymbol{P}_\text{2rad}^\alpha$ and $\boldsymbol{J}_{\mathcal{O}(G^4)}^{\alpha\beta}=\boldsymbol{J}_\text{1rad}^{\alpha\beta}+\boldsymbol{J}_{\text{2rad}}^{\alpha\beta}$ into two contributions involving either one (``1rad'') or two (``2rad'') radiation (on-shell) modes. The former, which we can cast as follows using the properties \eqref{eq:parity}, \eqref{eq:parityJ},
\begin{equation}\label{eq:1radPK}
	\boldsymbol{P}_\text{1rad}^\alpha
	=
	2
	\int_k 
	\operatorname{Re}
	\boldsymbol{K}^\alpha[\tilde{\mathcal{A}}_0, \tilde{\mathcal{B}}_{1O}^{(i)}+\tilde{\mathcal{B}}_{1E}]
	\,,
\end{equation}
and
\begin{equation}\label{eq:1radJO}
	\boldsymbol{J}_\text{1rad}^{\alpha\beta}
	=
	2
	\int_k 
	\operatorname{Im}
	\boldsymbol{O}^{\alpha\beta}[\tilde{\mathcal{A}}_0, \tilde{\mathcal{B}}_{1O}^{(i)}+\tilde{\mathcal{B}}_{1E}]
\,,
\end{equation}
lead to \emph{integer} PN corrections, while
the latter, 
\begin{equation}\label{eq:2radPK}
	\boldsymbol{P}_\text{2rad}^\alpha
	=
	\int_k 
	\left(
	2
	\operatorname{Re}
	\boldsymbol{K}^\alpha[\tilde{\mathcal{A}}_0, \tilde{\mathcal{B}}_{1O}^{(h)}]
	-
	\operatorname{Im}
	\boldsymbol{K}^\alpha[\tilde{\mathcal{A}}_0, \tilde{\mathcal{C}}]
	\right),
\end{equation}
and
\begin{equation}\label{eq:2radJO}
	\boldsymbol{J}_\text{2rad}^{\alpha\beta}
	=
	\int_k 
	\left(
	2
	\operatorname{Im}
	\boldsymbol{O}^{\alpha\beta}[\tilde{\mathcal{A}}_0, \tilde{\mathcal{B}}_{1O}^{(h)}]
	+
	\operatorname{Re}
	\boldsymbol{O}^{\alpha\beta}[\tilde{\mathcal{A}}_0, \tilde{\mathcal{C}}]
	\right),
\end{equation}
yield \emph{half-odd} PN corrections to the energy and angular momentum losses \cite{Georgoudis:2023eke,Heissenberg:2024umh}.

Since the kernel ingredients, such as $\mathcal{B}_{1}$, $\mathcal{C}$, are all real in momentum space, they obey
$\tilde{\mathcal{B}}_1^\ast = \tilde{\mathcal{B}}_1\big|_{b\to -b}$,
$\tilde{\mathcal{C}}^\ast = \tilde{\mathcal{C}}\big|_{b\to -b}$.
Performing then the same analysis based on the parity of the integrands under $b\cdot k\mapsto-b\cdot k$ as detailed above for the tree-level case, we find that the 1rad contributions follow the same pattern as their $\mathcal{O}(G^3)$ counterparts,
\begin{equation}\label{eq:1rad0components}
	b\cdot \boldsymbol{P}_\text{1rad} = 0\,,
	\qquad
	u_1\cdot \boldsymbol{J}_\text{1rad}  \cdot u_2= 0\,.
\end{equation} 
Instead, the ``bare'' factor of $i$ appearing in front of $\mathcal{C}$ (which follows from unitarity) induces a different pattern in the 2rad part, and the contributions of $\tilde{\mathcal{B}}_{1O}^{(h)}$ and $\tilde{\mathcal{C}}$ neatly separate among the components:
\begin{subequations}
	\begin{align}\label{eq:longitudinalP2rad}
		u_{1,2}\cdot \boldsymbol{P}_\text{2rad} &= 2u_{1,2}^\alpha
		\int_k
		\operatorname{Re}
		\boldsymbol{K}_\alpha[\tilde{\mathcal{A}}_0, \tilde{\mathcal{B}}_{1O}^{(h)}]\,,
		\\
		\label{eq:P2radC}
		b\cdot \boldsymbol{P}_\text{2rad} &= - b^\alpha \int_k 
		\operatorname{Im}
		\boldsymbol{K}_\alpha[\tilde{\mathcal{A}}_0, \tilde{\mathcal{C}}]\,,
	\end{align}
\end{subequations} 
and
\begin{subequations}
	\begin{align}\label{eq:longitudinalJ2rad}
		u_{1,2}\cdot \boldsymbol{J}_\text{2rad} \cdot b &= 2u_{1,2}^\alpha b^\beta
		\int_k
		\operatorname{Im}
		\boldsymbol{O}_{\alpha\beta}[\tilde{\mathcal{A}}_0, \tilde{\mathcal{B}}_{1O}^{(h)}]\,,
		\\
		\label{eq:J2radC}
		u_1\cdot \boldsymbol{J}_\text{2rad} \cdot u_2 &= u_1^\alpha u_2^\beta 
		\int_k 
		\operatorname{Re}
		\boldsymbol{O}_{\alpha\beta}[\tilde{\mathcal{A}}_0, \tilde{\mathcal{C}}]\,.
	\end{align}
\end{subequations} 
This analysis shows that the components $u_{1,2}\cdot \boldsymbol{P}_\text{2rad}$ and $u_{1,2}\cdot \boldsymbol{J}_\text{2rad} \cdot b$, which are the relevant ones to calculate the emitted energy and angular momentum in the center-of-mass frame, are fixed by the interference between the tree-level waveform and the RR piece of the one-loop waveform, $\mathcal{B}_{1O}^{(h)}$. Since, by Eq.~\eqref{eq:radreac(h)}, this RR piece is itself proportional to the tree-level waveform, Eqs.~\eqref{eq:longitudinalP2rad}, \eqref{eq:longitudinalJ2rad} can be entirely determined from the knowledge of the tree-level amplitude only.
 
\section{RR and energy loss}

Introducing the following notation for the longitudinal part of the 2rad energy-momentum loss,
\begin{equation}
	\boldsymbol{P}_{\parallel}^\alpha = (-u_1\cdot \boldsymbol{P}_\text{2rad})\check{u}_1^\alpha + (-u_2\cdot \boldsymbol{P}_\text{2rad}) \check{u}_2^\alpha\,,
\end{equation}
by Eq.~\eqref{eq:longitudinalP2rad} we have
\begin{equation}
	\boldsymbol{P}_{\parallel}^\alpha
	=
	2
	\int_k
	\operatorname{Re}
	\boldsymbol{K}^\alpha[\tilde{\mathcal{A}}_0, \tilde{\mathcal{B}}_{1O}^{(h)}]
\end{equation}
and, thanks to Eq.~\eqref{eq:radreac(h)}, this turns into
\begin{equation}\label{eq:PradExpanded}
	\boldsymbol{P}_{\parallel}^\alpha 
	=
	2\pi G \int_k (m_1\omega_1+m_2\omega_2) k^\alpha \tilde{\mathcal{A}}_0^\ast \tilde{\mathcal{A}}_0\,.
\end{equation}
This integral can be recast as the Fourier transform of a three-particle cut built from the product of two tree-level amplitudes (see Fig.~\ref{fig:3pc}), weighted by $(m_1\omega_1+m_2\omega_2) k^\alpha $, and thus calculated with the two-loop reverse-unitarity techniques developed in Ref.~\cite{Herrmann:2021lqe} (see also \cite{DiVecchia:2021bdo,Herrmann:2021tct}). The result is
\begin{equation}\label{eq:2radP}
	\boldsymbol{P}_{\parallel}^\alpha 
	\!=\!
	\frac{G^4m_1^2 m_2^2}{b^4}\!\left[
	m_1 ( \mathcal{E}^{(1)} \check{u}_1^\alpha 
	+
	\mathcal{E}^{(2)}  \check{u}_2^\alpha ) \!+\! (1\!\leftrightarrow\!2)
	\right]
\end{equation}
with 
\begin{equation}
	\mathcal{E}^{(i)} = \frac{f^{(i)}_1}{\sigma^2-1} + f^{(i)}_2 \,\frac{\operatorname{arccosh}\sigma}{(\sigma^2-1)^{3/2}} + f^{(i)}_3 \,\frac{(\operatorname{arccosh}\sigma)^2}{(\sigma^2-1)^{2}}
\end{equation}
for $i=1,2$ and the polynomials $f^{(i)}_{1,2,3}$ given in Appendix~\ref{app:Tables}. Note that \eqref{eq:2radP} can be more easily obtained by focusing on either term in \eqref{eq:PradExpanded}, and then using particle interchange symmetry to deduce the other one.

Eq.~\eqref{eq:2radP} is in perfect agreement with the results obtained in Ref.~\cite{Dlapa:2022lmu} from a worldline EFT and in Ref.~\cite{Damgaard:2023ttc} from an amplitude approach. Those references calculated the complete impulses, $Q_{1,2}^\alpha$ experienced by particles 1 and 2 up to $\mathcal{O}(G^4)$, including both 1rad and 2rad contributions, from which
the radiated energy-momentum can be deduced by the balance law $\boldsymbol{P}^{\alpha}= - Q_1^\alpha - Q_2^\alpha$.  

The result \eqref{eq:2radP} determines the full 2rad contribution to the radiated energy in the center-of-mass 
frame,\footnote{Note that $m_1 u^\mu_1+m_2 u^\mu_2=m_1 v^\mu_1+m_2 v^\mu_2 + \mathcal{O}(G^2)$.}
\begin{equation}\label{eq:E2radformal}
	E_\text{2rad} = -t \cdot \boldsymbol{P}_\text{2rad}\,,\qquad t^\mu = \frac{m_1 u^\mu_1+m_2 u^\mu_2}{E}\,.
\end{equation}
The first few terms in the expansion of this result for small $p_\infty = \sqrt{\sigma^2-1}$ read
\begin{equation}\label{eq:E2radPN}
	\begin{split}
	&E_\text{2rad}
	=
	\frac{G^4 M^5}{b^4} \nu^2p_{\infty }^2\Bigg[
	\frac{3136 }{45}
	+\left(\frac{1216}{105}-\frac{2272 \nu }{45}\right) p_{\infty
	}^2
	\\
	&+
	\left(\frac{117248}{1575}-\frac{8056 \nu }{1575}+\frac{1528 \nu ^2}{45}\right) p_{\infty }^4
	+\mathcal{O}(p_\infty^6)
	\Bigg]
	\end{split}
\end{equation}
with $M= m_1+m_2$ and $\nu=m_1 m_2/M^2$, again in agreement with the PN results obtained in \cite{Bini:2021gat,Bini:2022enm}.
Note that the PN expansion of $E_\text{1rad}$ provides the Newtonian (0PN) contribution at $\mathcal{O}(p_\infty^{-1})$ and only contributes odd powers of $p_\infty$, i.e.~integer PN corrections.
On the contrary, \eqref{eq:E2radPN} starts at $\mathcal{O}(p_\infty^2)$, thus at 1.5PN relative to the Newtonian order, and only involves even powers of $p_\infty$, i.e.~half-odd PN corrections.
Via \eqref{eq:E2radformal}, Eq.~\eqref{eq:2radP} provides the all-order resummation of such fractional PN contributions.

\section{RR and angular momentum loss}

Turning to the angular momentum loss, letting
\begin{equation}
		b^2
	\boldsymbol{J}_{\perp}^{\alpha\beta} 
	\!= \!
	(b\!\cdot\! \boldsymbol{J}_\text{2rad}\!\cdot\! \check u_1)\,
	u_1^{[\alpha} b_{\phantom{1}}^{\beta]} + 
	(b\!\cdot\! \boldsymbol{J}_\text{2rad}\!\cdot\! \check{u}_2)\,
	{u}_2^{[\alpha} b_{\phantom{2}}^{\beta]}
	\,,
\end{equation}
by Eq.~\eqref{eq:longitudinalJ2rad} we have
\begin{equation}
	\boldsymbol{J}_{\perp}^{\alpha\beta} 
	=
	2
	\int_k
	\operatorname{Im}
	\boldsymbol{O}^{\alpha\beta}[\tilde{\mathcal{A}}_0, \tilde{\mathcal{B}}_{1O}^{(h)}]\,.
\end{equation}
By Eq.~\eqref{eq:radreac(h)}, this turns 
into\footnote{The terms in which the derivatives act on $\omega_{1,2}$ cancel  due to $\overset{\leftrightarrow}{\partial}$.}
\begin{equation}\label{eq:JradExpanded}
\boldsymbol{J}_{\perp}^{\alpha\beta} 
	=
	-
	2 i \pi G \int_k (m_1\omega_1+m_2\omega_2) \,
	\boldsymbol{O}^{\alpha\beta}[\tilde{\mathcal{A}}_0, \tilde{\mathcal{A}}_0]
	\,.
\end{equation}
The integrand again reduces to an object quadratic in the tree-level waveform times an additional weighting function. The amplitude here does not appear multiplicatively, but techniques to reduce this type of integrals to the Fourier transform of a two-loop three-particle cut (Fig.~\ref{fig:3pc}) were developed in \cite{DiVecchia:2022piu,DiVecchia:2023frv} (see \cite{Heissenberg:2022tsn,Heissenberg:2023uvo} for related applications).
Although for the intermediate steps it is convenient to work in the translation frame $b_2^\alpha=0$, $b_1^\alpha = b^\alpha$, we present the result in the midpoint frame $b_1^\alpha = -b_2^\alpha = b^\alpha/2$, where particle-interchange symmetry becomes manifest (note that $b^\alpha \leftrightarrow -b^\alpha$ under $1\leftrightarrow2$),
\begin{equation}\label{eq:2radJ}
	\boldsymbol{J}_{\perp}^{\alpha\beta} 
	\!=\!
	\frac{G^4m_1^2 m_2^2}{b^3}\!\left[
	m_1 ( \mathcal{F}^{(1)} b_{\phantom{1}}^{[\alpha}u_1^{\beta]} + \mathcal{F}^{(2)} b_{\phantom{2}}^{[\alpha}u_2^{\beta]} ) \!+\! (1\!\leftrightarrow\!2)
	\right]
\end{equation}
where 
\begin{equation}
	\mathcal{F}^{(i)} = \frac{g^{(i)}_1}{(\sigma^2-1)^2} + g^{(i)}_2 \,\frac{\operatorname{arccosh}\sigma}{(\sigma^2-1)^{5/2}} + g^{(i)}_3 \,\frac{(\operatorname{arccosh}\sigma)^2}{(\sigma^2-1)^{3}}
\end{equation}
and the polynomials $g^{(i)}_{1,2,3}$ are given in Appendix~\ref{app:Tables}.
One can freely translate the result along $b^\alpha$ by means of \eqref{eq:translation} thanks to the explicit expression \eqref{eq:2radP} for $\boldsymbol{P}^\alpha_{\parallel}$. 
Like for $\boldsymbol{P}_{\parallel}^{\alpha}$, it is easier to first calculate one of the two terms in \eqref{eq:JradExpanded} and then obtain the other one by $1\leftrightarrow2$.

As a cross check, using the probe-limit, $m_1\to0$, waveform evaluated up to 10PN in the frame where $m_2$ is at rest, $b^\alpha_2=0$ and $u_1^\alpha=(\sigma, 0, p_\infty,0)$ \cite{toapCHRR}, we computed the radiated angular momentum in this frame for small $p_\infty$,
\begin{equation}\label{eq:PNdataJ}
	\begin{split}
		&\boldsymbol{J}_\text{2rad}^{\text{probe}}=
		\frac{G^4 m_1^2 m_2^3}{b^3} \Big[
		\frac{448 p_{\infty }}{5}
		+\frac{1184 p_{\infty }^3}{21}
		-\frac{13736 p_{\infty }^5}{315}
		\\
		&
		+\frac{724868 p_{\infty }^7}{17325}
		-\frac{15578279 p_{\infty}^9}{450450}
		+\frac{20316617 p_{\infty }^{11}}{700700}
		\\
		&-\frac{3525071503 p_{\infty }^{13}}{142942800}
		+\frac{1039071734251 p_{\infty
			}^{15}}{48886437600}
		\\
		&-\frac{14500043393593 p_{\infty }^{17}}{782183001600}
		+\frac{11996977412779 p_{\infty }^{19}}{734294246400}
		\\
		&-\frac{23005919863020091 p_{\infty}^{21}}{1583138395238400} \Big]
		+O\left(p_{\infty }^{23}\right) 
	\end{split}
\end{equation}
in 
agreement\footnote{One could also use the PN data \eqref{eq:PNdataJ} to bootstrap the full result \eqref{eq:2radJ} following the strategy adopted in \cite{Manohar:2022dea}.} 
with the prediction obtained from \eqref{eq:2radJ} after transforming it back to $b_2^\alpha=0$ by means of a translation \eqref{eq:translation}  with $a_\text{rest\,2}^\alpha = +\frac{1}{2}\,b^\alpha$, as $m_1\to0$,
\begin{equation}
	- p_\infty \frac{b_{\alpha}}{b}\left(\boldsymbol{J}_\text{2rad}^{\alpha\beta} + \tfrac{1}{2}\,b^{[\alpha}_{\phantom{2}} \boldsymbol{P}_\text{2rad}^{\beta]}\right) \check{u}_{1\beta}
	\sim \boldsymbol{J}_\text{2rad}^{\text{probe}}\,.
\end{equation}
Note that only the $\boldsymbol{J}_{\perp}^{\alpha\beta}$ and $\boldsymbol{P}_{\parallel}^{\alpha}$ contribute nontrivially to this combination. 

To transform instead \eqref{eq:2radJ} to the center-of-mass translation frame, $E_1 b_1^\alpha + E_2 b_2^\alpha=0$ where $E_{1,2}=-m_{1,2} \,t\cdot u_{1,2}$, one needs to perform a translation \eqref{eq:translation} by 
\begin{equation}
a^\alpha_\text{CM} = \frac{m_2^2-m_1^2}{2E^2}\,b^\alpha\,.
\end{equation}
Thus, the 2rad contribution to the radiated angular momentum defined in the center-of-mass frame is given by
\begin{equation}\label{eq:2radJCM}
	\boldsymbol{J}_\text{2rad} = \frac{b_\alpha}{b} \left( \boldsymbol{J}_\text{2rad}^{\alpha\beta}   + a_\text{CM}^{[\alpha} \boldsymbol{P}_\text{2rad}^{\beta]}\right) \frac{p_\beta}{p}\,,
\end{equation}
where $p^\alpha$
is the spatial momentum of particle 1 in the center-of-mass frame \eqref{eq:pCM} and $p=m_1 m_2\sqrt{\sigma^2-1}/E$ is its magnitude.
Let us note that, as anticipated, 
only the perpendicular $\boldsymbol{J}_{\perp}^{\alpha\beta}$ and longitudinal $\boldsymbol{P}_{\parallel}^{\alpha}$ components enter the combination \eqref{eq:2radJCM} that defines the radiated angular momentum in the center-of-mass frame.
In the small-velocity limit, we find
\begin{equation}\label{eq:J2radPN}
	\begin{split}
		&\boldsymbol{J}_\text{2rad}
		=\frac{G^4 M^5}{b^3}\nu^2 p_\infty
		\Bigg[
		\frac{448}{5}
		+
		\left(\frac{1184}{21}-\frac{45664 \nu }{315}\right) p_{\infty}^2
		\\
		&
		+	\left(\frac{10648 \nu ^2}{63}-\frac{2872 \nu }{63}-\frac{13736}{315}\right) p_{\infty }^4
		+\mathcal{O}(p_\infty^6)
		\Bigg],
	\end{split}
\end{equation}
where the first line reproduces the last line in Eq.~(5.17) of \cite{Heissenberg:2024umh}. (We also find agreement with the third line of Eq.~(5.31) of that Ref. by projecting on the appropriate component.) For reference, we recall that the Newtonian correction to the angular momentum loss at $\mathcal{O}(G^4)$ scales like $p_\infty^{-2}$, and arises from the 1rad contributions which are responsible for all integer-PN terms. 

Finally, in addition to the radiative loss, we need to account also for the static contribution already discussed and calculated in \cite{Heissenberg:2024umh,DiVecchia:2022owy}. By (5.41) of \cite{Heissenberg:2024umh},  this reads
\begin{equation}
	\mathcal{J}_\text{2rad} = \frac{G^2 p}{2b} Q_\text{1PM}^2 \mathcal{I}(\sigma)^2
\end{equation}
with $Q_\text{1PM}$ and $\mathcal{I}(\sigma)$ given in \eqref{eq:Q1PM}, \eqref{eq:I}. The complete expression for the total angular momentum lost by the system that resums all half-off PN orders is thus
\begin{equation}
		{J}_\text{2rad}  = 	\boldsymbol{J}_\text{2rad}  + 	\mathcal{J}_\text{2rad} \,,
\end{equation}
whose first few orders in the PN expansion read
\begin{align}
	\nonumber
		&{J}_\text{2rad}
		=\frac{G^4 M^5}{b^3}\nu^2 p_\infty
		\Bigg[
		\frac{448}{5}
		+\left(\frac{1184}{21}-\frac{220256 \nu }{1575}\right) p_{\infty }^2
		\\
		\label{eq:2radJfinal}
		&
		+	\!	\left(\frac{262168 \nu ^2}{1575}-\frac{46456 \nu }{1575}-\frac{13736}{315}\right) p_{\infty }^4
		+\mathcal{O}(p_\infty^6)
		\Bigg].
\end{align}
The first line of \eqref{eq:2radJfinal} matches the 1.5PN and 2.5PN corrections obtained in \cite{Bini:2021gat,Bini:2022enm} (more recently reproduced in \cite{Heissenberg:2024umh}).
We provide $\boldsymbol{P}_{\parallel}^\alpha$, $E_\text{2rad}$, $\boldsymbol{J}_{\perp}^{\alpha\beta}$, $\mathcal{J}_{\text{2rad}}$ and $J_{\text{2rad}}$ in computer-friendly format in the ancillary file \texttt{2rad-anc.m}.

\section{Discussion and outlook}

In this work, we pointed out that, at $\mathcal{O}(G^4)$, the  longitudinal components of $\boldsymbol{P}^{\alpha}_{\text{2rad}}$ and the transverse ones of $\boldsymbol{J}_{\text{2rad}}^{\alpha\beta}$ can be reduced to two-loop integrals, contrary to the three-loop order that naive power counting would suggest. These are precisely the components relevant to the radiated energy and angular momentum as defined in the center-of-mass frame. We leveraged this simplification to calculate the 2rad contribution to the total angular momentum loss, obtaining a new expression which resums all half-odd PN contributions to this observable at $\mathcal{O}(G^4)$.

Naturally it will be important and interesting to complete the present analysis by explicitly evaluating the remaining terms in Eqs.~\eqref{eq:1radPK}, \eqref{eq:1radJO}, \eqref{eq:2radPK}, \eqref{eq:2radJO}.
These include the 2rad parts involving $\mathcal{\tilde{C}}$ in \eqref{eq:P2radC}, \eqref{eq:J2radC}, which would serve as a cross-check of the results in \cite{Dlapa:2022lmu,Damgaard:2023ttc} for the radiated spatial momentum component $b\cdot \boldsymbol{P}_\text{2rad}$ and give new expressions for the radiated mass dipole moment or ``boost charge'' component $u_1\cdot  \boldsymbol{J}_\text{2rad}\cdot u_2$. The latter quantity, as opposed to the conventional angular momentum or ``rotation charge'' considered here, is however sensitive to the arbitrariness under time translations induced by  the infrared divergences in the Compton cuts, see \cite{Heissenberg:2024umh}, as well as to the supertranslation contribution discussed in Refs.~\cite{Veneziano:2022zwh,Georgoudis:2023eke,Georgoudis:2024pdz,Bini:2024rsy}. 
Its physical meaning thus appears to be quite subtle.
Perhaps more interesting are the $b \cdot  \boldsymbol{J}_\text{1rad}\cdot u_{1,2}$ components, which would provide a new result resumming also the integer PN corrections to the $\mathcal{O}(G^4)$ angular momentum loss. Of course, $2\int_k  \operatorname{Re} \boldsymbol{K}^\alpha[\tilde{\mathcal{A}}_0, \tilde{\mathcal{B}}_{1E}]$ and $2
\int_k 
\operatorname{Im}
\boldsymbol{O}^{\alpha\beta}[\tilde{\mathcal{A}}_0,\tilde{\mathcal{B}}_{1E}]$ represent the more challenging part of the calculations, while, thanks to \eqref{eq:radreac(h)},
\begin{equation}\label{eq:finalremark}
\begin{split}
	2\int_k  \operatorname{Re} \boldsymbol{K}^\alpha[\tilde{\mathcal{A}}_0, \tilde{\mathcal{B}}_{1O}^{(i)}] &= -\frac{\sigma(\sigma^2-\frac{3}{2})}{(\sigma^2-1)^{3/2}} \, \boldsymbol{P}_\parallel^\alpha\,,\\
	2
	\int_k 
	\operatorname{Im}
	\boldsymbol{O}^{\alpha\beta}[\tilde{\mathcal{A}}_0,\tilde{\mathcal{B}}_{1O}^{(i)}]
	&=
	-\frac{\sigma(\sigma^2-\frac{3}{2})}{(\sigma^2-1)^{3/2}} \, \boldsymbol{J}_\perp^{\alpha\beta}\,,
\end{split}
\end{equation}
which are thus proportional to the results \eqref{eq:2radP}, \eqref{eq:2radJ} above.
Since $\mathcal{B}_{1E}$ starts at 1PN, Eqs.~\eqref{eq:finalremark} fix the Newtonian (0PN) contributions to be $\frac{1}{2}$ times the 1.5PN ones in \eqref{eq:E2radPN}, \eqref{eq:J2radPN} in agreement with known data points \cite{Bini:2021gat,Bini:2022enm,Heissenberg:2024umh}.

Another interesting direction consists in calculating the individual mechanical contributions of particles 1 and 2 to the angular momentum loss, as done in \cite{DiVecchia:2022piu} at $\mathcal{O}(G^3)$, which would allow one to explicitly check the total balance law, as discussed for $Q_1$, $Q_2$ in \cite{Dlapa:2022lmu,Bini:2022enm}. Further natural generalizations consist in taking into account tidal and spin effects \cite{Riva:2022fru,Mougiakakos:2022sic,Heissenberg:2022tsn,Heissenberg:2023uvo,Jakobsen:2023hig}, leveraging in particular the spinning waveforms obtained in \cite{Jakobsen:2021lvp,Riva:2022fru,DeAngelis:2023lvf,Brandhuber:2023hhl,Aoude:2023dui,Bohnenblust:2023qmy}, or additional massless fields such as those appearing in $\mathcal{N}=8$ supergravity, which can also serve as a simpler testing ground for more challenging calculations \cite{Bern:2020gjj,Parra-Martinez:2020dzs,DiVecchia:2021bdo,Herrmann:2021tct}. An intriguing open issue also concerns the high-energy behavior of radiated quantities, whose ill-behaved large-$\sigma$ expansion points to a breakdown of the conventional PM approximation, thus requiring to develop novel tools for its investigation \cite{DiVecchia:2022nna,Alessio:2024onn,Haring:2024wyz,Rothstein:2024nlq}.

\section*{Acknowledgments.}
I would like to thank P.~H. Damgaard, S.~Foffa, L.~Plant\'e, M.~Porrati, S.~Speziale for discussions, and P.~Di Vecchia, R.~Russo  also for helpful comments on a preliminary version of this work.

\appendix

\section{Notation and conventions}

We collect a few equations concerning notation and conventions in this appendix.
The Fourier transform from momentum to impact-parameter space reads
\begin{equation}\label{eq:tildeW}
	\tilde{W}^{\mu\nu}(k)
	=
	\int_{q_1,q_2}
	e^{ib_1\cdot q_1+ ib_2\cdot q_2}W^{\mu\nu}(q_1,q_2)
\end{equation}
with (letting $p_{1,2}$ denote the incoming momenta)
\begin{equation}
	\int_{q_1,q_2}\!\!\!\!\!
	=\!\! \int\!\!  \frac{d^Dq_1d^Dq_2}{(2\pi)^{D-2}}	\delta(2p_1\cdot q_1) \delta(2p_2\cdot q_2)
	\delta^{(D)}(q_1+q_2+k)\,.
\end{equation}
The phase-space integration over the emitted graviton momentum is given by
\begin{equation}\label{eq:intk}
	\int_k = \int \frac{d^Dk}{(2\pi)^{D}}\,2\pi\theta(k^0)\delta(k^2)\,.
\end{equation}

Let us also quote a more detailed expression for the regulated Compton cuts $\mathcal{C}$ appearing in Eq.~\eqref{eq:kernelWreg},
\begin{equation}\label{eq:ComptonCutsExplicit}
	\frac{i}{2}\,\mathcal{C}^{\mu\nu} 
	= 2i GE \omega \log\frac{\omega}{\mu_\text{IR}} \mathcal{A}^{\mu\nu}_0 + (C^{\text{reg}})^{\mu\nu}\,,
\end{equation}
which facilitates the comparison with \cite{Georgoudis:2023eke,Georgoudis:2024pdz}.
We recall that
\begin{equation}\label{eq:Q1PM}
	Q_\text{1PM}  = \frac{4 G m_1 m_2 (\sigma^2-\frac{1}{2})}{b\sqrt{\sigma^2-1}}
\end{equation}
is the leading-order impulse and 
\begin{equation}\label{eq:I}
	\mathcal{I}(\sigma) = \frac{2 \sigma ^2}{\sigma ^2-1}-\frac{16}{3}+\frac{2 \left(2 \sigma ^2-3\right) \sigma\, \text{arccosh}\sigma}{\left(\sigma ^2-1\right)^{3/2}}
\end{equation}
the RR function first introduced in Ref.~\cite{Damour:2020tta}. 

The spatial momentum of particle 1 in the center-of-mass frame takes the form
\begin{equation}\label{eq:pCM}
	p^\alpha = \frac{m_1 m_2}{E^2}\left(
	m_1 (\sigma u_1^\alpha - u_2^\alpha) - m_2  (\sigma u_2^\alpha - u_1^\alpha) 
	\right).
\end{equation}

\section{Polynomials entering $E_\text{2rad}$, $J_\text{2rad}$}
\label{app:Tables}

We present here the explicit expressions of the polynomials appearing in Eq.~\eqref{eq:2radP},
			\begin{align*}
				f^{(1)}_1 & =
				\frac{16}{45} \left(12 \sigma ^6-1484 \sigma ^4+4927 \sigma ^2+1720\right)
				\\
				f^{(1)}_2 &= \frac{16}{5} \sigma  \left(16 \sigma ^6+204 \sigma ^4-496 \sigma ^2-869\right)
				\\
				f^{(1)}_3 &= -32 \left(8 \sigma ^6-6 \sigma ^4-51 \sigma ^2-8\right)
				\\
				f^{(2)}_1 &= \frac{16}{3} \sigma  \left(64 \sigma ^4-130 \sigma ^2-411\right)
				\\
				f^{(2)}_2 &=-\frac{16}{3} \left(64 \sigma ^6+20 \sigma ^4-868 \sigma ^2-173\right)
				\\
				f^{(2)}_3 &= 64 \sigma  \left(\sigma ^2+4\right) \left(2 \sigma ^4-5 \sigma ^2-5\right),
			\end{align*}
and in Eq.~\eqref{eq:2radJ},
			\begin{align*}
				g^{(1)}_1 &=\frac{8}{45} \left(972 \sigma ^6+1456 \sigma ^4+10177 \sigma ^2-320\right)
				\\
				g^{(1)}_2 &=-\frac{8}{5} \sigma  \left(64 \sigma ^6+76 \sigma ^4+2336 \sigma ^2+259\right)
				\\
				g^{(1)}_3 &= 16 \left(-4 \sigma ^6+100 \sigma ^4+39 \sigma ^2+2\right)
				\\
				g^{(2)}_1 & =
				\frac{8}{45} \sigma  \left(564 \sigma ^6-2732 \sigma ^4+11347 \sigma ^2+3061\right)
				\\
				g^{(2)}_2 &= -\frac{8}{15} \left(48 \sigma ^8-764 \sigma ^6+3908 \sigma ^4+4829 \sigma ^2+169\right)
				\\
				g^{(2)}_3 &= 16 \sigma  \left(-12 \sigma ^6+36 \sigma ^4+95 \sigma ^2+18\right).
			\end{align*}

\providecommand{\href}[2]{#2}\begingroup\raggedright\endgroup

\end{document}